\RequirePackage{lineno}
\documentclass[preprint,preprintnumbers,amsmath,amssymb]{revtex4}

\usepackage{graphicx}
\usepackage[bookmarks=false]{hyperref}
\usepackage{amsmath}
\usepackage{amssymb}
\usepackage{bm}
\usepackage{amsthm}
\usepackage{amsfonts}
\usepackage{latexsym}
\usepackage{wrapfig}
\usepackage[usenames,dvipsnames]{color}
\usepackage[protrusion=true,expansion=true,final]{microtype}
\usepackage{ifthen}
\usepackage{tikz}\usetikzlibrary{shapes,arrows,positioning}
\usepackage{color,soul}
\usepackage{lineno}
\usepackage{multirow}
\usepackage{booktabs}
\usepackage[normalem]{ulem}
\useunder{\uline}{\ul}{}

\begin{document}

\title{Surface- and strain-tuning of the optical dielectric function in epitaxially grown CaMnO$_3$}
\author{Dominic Imbrenda$^1$}
\author{Dongyue Yang$^2$}
\author{Hongwei Wang$^2$}
\author{Andrew R. Akbashev$^3$}
\author{Leila Kasaei$^2$}
\author{Bruce A. Davidson$^2$}
\author{Xifan Wu$^2$}
\author{Xiaoxing Xi$^2$}
\author{Jonathan E. Spanier$^{1,3,4}$}
\email[ ]{email: spanier@drexel.edu}
\affiliation{$^1$Department of Electrical \& Computer Engineering,\!
	Drexel University,\! Philadelphia,\! PA 19104,\! USA}%
\affiliation{$^2$Department of Physics,\!
	Temple University,\! Philadelphia,\! PA 19104,\! USA}%
\affiliation{$^3$Department of Materials Science \& Engineering,\!
	Drexel University,\! Philadelphia,\! PA 19104,\! USA}%
\affiliation{$^4$Department of Physics,\!
	Drexel University,\! Philadelphia,\! PA 19104,\! USA}%
%
\date{\today}
%

\begin{abstract}
\noindent{We report a strong thickness dependence of the complex frequency-dependent optical dielectric function $\widetilde{\epsilon}(\omega)$ over a spectral range from 1.24 to 5 eV in epitaxial CaMnO$_3$(001) thin films on SrTiO$_3$(001), LaAlO$_3$(001), and SrLaAlO$_4$(001). A doubling of the peak value of the imaginary part of $\widetilde{\epsilon}(\omega)$ and spectral shifts of 0.5 eV for a given magnitude of absorption are observed. On the basis of experimental analyses and first-principles density functional theory calculations, contributions from both surface states and epitaxial strain to the optical dielectric function of CaMnO$_3$ are seen. Its evolution with thickness from 4 to 63 nm has several regimes.  In the thinnest, strain-coherent films, the response is characterized by a significant contribution from the free surface that dominates strain effects. However, at intermediate and larger thicknesses approaching the bulk-like film, strain coherence and partial strain relaxation coexist and influence $\widetilde{\epsilon}(\omega)$.}
\end{abstract}
\maketitle

Strain engineering has long been used as an effective way to tune electronic, magnetic and optical properties of oxide thin films~\cite{spaldin_strain_1, bhattacharjee2009engineering}. The lattice distortions imposed by epitaxial strain can introduce dramatic changes in the properties of thin-film materials, e.g. allowing strong ferroelectric ordering in quantum paraelectrics~\cite{haeni2004room}, manipulation of transition temperatures in ferroelectrics~\cite{strain_FE_1}, tuning of magnetic and metal-insulator transitions in mixed-valence perovskite oxides~\cite{strain_MIT_1,strain_MIT_2}, and controlling the volume of the magnetic phase in magnetically inhomogeneous media ~\cite{phase_separ_1}. Owing to the direct relationship between the electronic structure and optical properties, epitaxial strain strongly influences dielectric constants, refractive indices and, ultimately, the bandgap of a thin film material ~\cite{singh2014strain,scafetta2014band,liu2013strain,scafetta2013optical}. However, such studies for oxides are still scarce, and the roles of chemistry, structure, native defects (oxygen vacancies), film thickness and surface effects (termination, admolecules, structural reconstruction) have yet to be clarified.

In the bulk, electronic structure and optical properties of CaMnO$_3$ (CMO) are well studied~\cite{jung1997determination,loshkareva2004electronic,molinari2014structural}, yet little attention has been given to the optical properties of epitaxial CMO thin films. CMO has long been known as an archetypal mixed-valence manganite that exhibits colossal magnetoresistance (CMR). However, strain-induced multiferroicity has also been predicted for CMO ~\cite{bhattacharjee2009engineering} with incipient ferroelectricity being later confirmed in tensile-strained films ~\cite{CMO_FE}. Similar to other perovskite oxides, CMO can easily accommodate oxygen vacancies that can be introduced allowing the material to demonstrate modest electrocatalytic activity ~\cite{CMO_catalyst_1, CMO_catalyst_2}. Still, little is known about the changes in the electronic structure in thin and ultrathin CMO films despite a certain surge of interest to strain-mediated effects on the optical properties of perovskite oxides ~\cite{dejneka2010tensile, liu2013strain, scafetta2014band, singh2014strain, roy2012effects,chernova2015strain,Choi2014LASTO}. In addition, because surface reconstruction and termination become crucial in a few unit cell thick perovskite oxides, ultrathin CMO films are expected to demonstrate optical properties that are distinct from those of thicker films ~\cite{saldana2015structural}.

Here we report pronounced thickness dependence of the complex frequency-dependent optical dielectric function $\widetilde{\epsilon}(\omega)$ in epitaxial CMO thin films. Using spectroscopic ellipsometry we perform detailed characterization of optical properties of CMO thin films, and we employ first-principles density functional theory (DFT) calculations to determine the contribution both from the surface states and epitaxial strain to the optical dielectric function of CMO.
 
Epitaxial CMO thin films were grown by pulsed laser deposition (PLD) using a KrF excimer laser ($\lambda$ = 248 nm) on single-crystal (001)-oriented SrTiO$_3$ (STO), LaAlO$_3$ (LAO), and SrLaAlO$_4$ (SLAO) purchased from Crystec ~\cite{crystec}. A laser repetition rate of 2.11 Hz and a laser energy density of 2.0 J/cm$^2$ were used. The substrate temperature was 650$^\circ$C and the background pressure was around 8.6$\times$10$^{-6}$ Torr. Films were deposited in oxygen environment at 30 mTorr and after the deposition, the sample was cooled to room temperature within an oxygen environment around 300 Torr to reduce/eliminate oxygen vacancies in the films. The thicknesses of the films were measured by X-ray reflectivity (XRR) (Fig. 1(d)) and the film deposition rate was determined to be approximately 0.74 \AA/pulse. Samples ranging from 4.1 to 63 nm in thickness were each grown on SrTiO$_3$(001), and films ranging in thickness from 4.3 to 10 nm were grown on LAO(001) and SLAO(001). Grazing incidence X-ray diffraction (GIXRD) ~\cite{Lee2011} (Fig. 1(a)) and reciprocal space mapping (RSM) were used to determine film lattice strain states by measuring the film in-plane lattice constant for thin ($<$20 nm) and thick ($>$20 nm) films respectively. As can be seen in Fig. 1(a), the thinnest film (4.1 nm) shows no substrate peak broadening and no extra peak, indicating the fully strained status of the film. In the 7.1 nm sample curve, the shoulder of the STO peak is evidence that the film starts to relax but is still mainly strained to the substrate. In the 10.4 nm sample curve, no shoulder at the STO peak position and the bump at the CMO(002) position shows a mainly relaxed state. X-ray diffraction (XRD) (Fig. 1(b)) was used to ensure the films were single phase crystals with no secondary phases and to measure out-of-plane lattice parameter. Atomic force microscopy (AFM) (Fig. 1(c)) was used to confirm film surface quality.

Five CMO film samples of different thicknesses ranging from 4.1 nm to 62.9 nm were grown on STO, and two additional films of $\sim$10 nm and $\sim$4 nm were grown, each on LAO and SLAO, to probe and disentangle the effects of thickness, strain and strain relaxation on $\widetilde{\epsilon}(\omega)$ of CMO. The bulk in-plane lattice parameter of STO ($a_\text{STO}$ = 3.905 \AA) is larger than that for both LAO ($a_\text{LAO}$ = 3.790 \AA) and SLAO ($a_\text{SLAO}$ = 3.754 \AA) and largest compared to that for bulk CMO ($a_\text{CMO}$ = 3.72 \AA). As STO has the largest in-plane lattice parameter compared to bulk CMO, the in-plane CMO film strain coherency in CMO/STO persists to smaller thicknesses than for CMO films on LAO or SLAO. A summary of the parameters for each film, including substrate, thickness, measured in-plane lattice parameter $a_\parallel$, the corresponding in-plane strain, and surface roughness, is shown in Table I.

Variable-angle spectroscopic ellipsometry (VASE) was performed at room temperature in ambient atmosphere with an electronically controlled rotating compensator and Glan-Taylor polarizers (J.A. Woollam, M2000). Measurements were performed at multiple angles between 65-75$^\circ$ and in the spectral range of 247 to 1000 nm with a resolution of 1.6 nm. Measurement of the components of linearly polarized reflectivity at each selected wavelength were used to obtain the ellipsometric parameters
$\Psi$ and $\Delta$. To determine $\widetilde{\epsilon}(\omega)$ for CMO we assume a four-layer optical medium comprised of a homogeneous isotropic film layer on a semi-infinite bulk, incorporating surface roughness, in vacuum. The surface roughness layer is modeled using the Bruggeman effective medium approximation using a 50\% film and 50\% void at the surface \cite{tompkins1999spectroscopic} with thickness approximately equal to the rms roughness for each sample obtained from XRR. Our model accounting for surface effects is valid as long as the surface layers are thin compared to the bulk and have a refractive index lower than that of the bulk \cite{nelson2012dielectric}. Optical dielectric functions of each substrate from the same batch as those used for film growth were also determined, assuming a semi-infinite half-space in vacuum, including surface roughness, and incorporated as the bulk layer in the model.

To fit the spectral dependence and calculate the complex index of refraction of each sample a combination of Gaussian curves and Lorentz oscillators~\cite{tompkins1999spectroscopic} were used to model the experimental data for the CMO films. Both of these functions are Kramers-Kronig (K-K) consistent which ensures causality. Experimental data for the STO and SLAO substrates were modeled using Gaussian curves and Lorentz oscillators however, as the spectral range for measurement in our setup is below the bandgap of LAO~\cite{lim2002dielectric} a Cauchy function was used to model its optical response. The number and type of oscillators used was sample dependent.  Regression analysis using the Levenberg-Marquardt algorithm was performed until the weighted mean squared error (MSE) between the calculated and experimental data was minimized. Thickness of the film and surface roughness were determined from XRR and held constant in the model until a satisfactory fit was achieved. After obtaining an acceptable MSE the thickness and surface roughness were made free parameters to ensure good agreement of thickness between the model and XRR data. VASE- and XRR- determined thicknesses, along with the MSE for each film, are given in Table I. 

$\widetilde{\epsilon}(\omega)$ depends sensitively on film thickness (Fig. 2). For films grown on STO, increasing thickness is accompanied by progressive strain relaxation and evolution of $\widetilde{\epsilon}(\omega)$. For our thickest, bulk-like film (62.9 nm), $\widetilde{\epsilon}(\omega)$ is consistent with previously published data for bulk CMO crystals ~\cite{loshkareva2004electronic,nomerovannaya2006ellipsometric}. For thinner films and increasing thickness fraction of strain coherence, to fully strain-coherent (about 4 and 6 nm films), there is reduction by as much as 50\% of both the real and imaginary parts of $\widetilde{\epsilon}(\omega)$. The linear portion of the optical absorption $\alpha(\omega)$ = 4$\pi k$/$\lambda$ shifts by as much as 0.5 eV with film thickness in the range of 2 $<$ $\hbar\omega$ $<$ 2.5 eV (Fig. 2(a)). This is due to the shift in energy of the first peak in the imaginary component ($\epsilon_{2}$) of $\widetilde{\epsilon}(\omega)$ and a reduction in transition strength as we discuss below.

To further understand the origin of the thickness-dependence of $\widetilde{\epsilon}(\omega)$ we analyzed CMO films of comparable thicknesses, deposited on LAO and SLAO. The in-plane lattice parameter of SLAO is closest to CMO and the films deposited on SLAO are under a smaller in-plane strain compared to films on STO (Table I). The peak value $\epsilon_{2}$ is larger for a film deposited on SLAO than for a film of comparable thickness deposited on STO (Fig. 3). LAO has an in-plane lattice constant intermediate to SLAO and STO and films deposited on LAO are less strained than those deposited on STO but are under more strain than those on SLAO. This results in a peak value of $\epsilon_{2}$ intermediate to that of films of similar thickness deposited on SLAO and STO (Fig. 3). GIXRD data (Fig. 1(a)) indicate that both SLAO films and both LAO films are fully strained to the substrate, yet we still observe an evolution of $\epsilon_{2}$. If strain was the dominant effect, we would expect $\epsilon_{2}$ to be the same for a given strain state.

First-principles calculations were performed by using DFT as implemented in the Vienna Ab Initio Simulation Package (VASP)\cite{kresse93, kresse96}. The electron exchange and correlation were approximated by the generalized gradient approximation revised for solid (PBEsol)\cite{perdew08} and we adopted an effective on-site Coulomb repulsion $U-J = 3.0~{\rm eV}$ for the $d$ orbitals of Mn atoms~\cite{Jiawang12}. The Kohn-Sham equations were expanded by plane-wave bases truncated at a cutoff energy of 500 eV. A $k$-point mesh of 6 $\times$ 6 $\times$ 1 was used in both the calculation of the (CMO)$_{4}$(STO)$_{4}$(CMO)$_{4}$ supercell and the calculation of the surface between this supercell and vacuum ((CMO)$_4$(STO)$_4$(CMO)$_4$-$surface$), in which vacuum is approximated by 10 {\AA} distance along the [001] direction within the periodic boundary condition. For the studies of bulk CMO strain coherent with STO, a denser $k$-point mesh of 6 $\times$ 6 $\times$ 4 was used. The lattice constant $c$ along the [001] direction as well as all the atomic positions are fully relaxed with the remaining Hellman-Feynman force being less than 1 m{\rm eV}/{\AA}. In all the \textit{surface} theoretical models, we consider the surface relaxation by including the vacuum in the model. The vacuum spans in the space for about 10 {\AA} above the surface of the CMO under the periodic boundary condition. In order to determine the stability of different surface terminations, we performed the density functional theory total energy minimizations for both MnO$_2$ terminated and CaO terminated surface models, in which all the atomic positions are allowed to relax. It was found that the MnO$_2$ terminated surface is always more energetically stable than CaO in all the slabs. As a result, the MnO$_2$ terminated surface models are used in our theoretical simulations.

Notably in the films of $\sim$5 nm, equilibrium surface atomic structure and electronic wavefunction mixing of the substrate with the bottom several unit cells of CMO can contribute to $\widetilde{\epsilon}(\omega)$, in addition to strain. DFT calculations were performed, considering several candidate structures. To provide further insight into the primary origin of the thickness-dependence we calculated $\widetilde{\epsilon}(\omega)$ in the in-plane direction for four and six unit-cell supercells of CMO. We first omitted contributions from strain from a substrate to determine the effect of energy-minimized surface truncation on $\widetilde{\epsilon}(\omega)$. Remarkably, the effect is very pronounced for a four unit-cell ($\sim$3.2 nm-thick) film, and considerably weaker for the 6 unit-cell ($\sim$4.8 nm-thick) film (Fig. 4(a)). Incorporation of epitaxial strain (Fig. 4(b)) by introducing STO into a four unit-cell supercell, (CMO)$_4$(STO)$_4$(CMO)$_4$, with no free surface, reduces $\epsilon_{2}$, but less than the reduction with a (CMO)$_4$ supercell with one free surface. Significantly, for the 4-layer CMO film with a free, vacuum-terminated surface, the effect of including substrate strain i.e., (CMO)$_4$(STO)$_4$(CMO)$_4$-$surface$ is relatively small (Fig. 4(b)). The first peak in the imaginary part of in-plane dielectric function is barely changed; while the second peak is slightly higher in the computed $\epsilon_{2}$ of (CMO)$_4$(STO)$_4$(CMO)$_4$-$surface$ than that of the 4-layer CMO film with a free, vacuum-terminated surface. This is due to the additional spectra signals from the STO component. Close inspection on the transition matrices involved in the calculation of dielectric function reveals that the first peak of the imaginary part of the in-plane dielectric function can be assigned to the hopping processes between equatorial oxygen \textit{p} orbitals and the e$_g$ electrons of $d_{x^{2}-y^{2}}$ character. At the surface, a MnO$_2$ terminated slab is always found to be more energetically favorable than the SrO terminated one. As result, the Mn atom at MnO$_2$ terminated surface undergoes an abrupt reconstruction due to the missing of one apical oxygen atoms. Such a surface reconstruction in turn largely increases the $d_{x^{2}-y^{2}}$ levels originating from the shortened Mn-O bond length and therefore increases the Coulombic energy. The phonon frequencies due to the hopping processes between occupied equatorial oxygen \textit{p} and the non-occupied e$_g$ ($d_{x^{2}-y^{2}}$) electrons also shift to higher energy. As a result, the surface reconstruction shifts the first peak of $\epsilon_{2}$ towards higher energy with a reduced magnitude. This optical effect from surface reconstruction will be more pronounced when the film goes to the thin limit as seen in both experiment and theory in Figs. 3 and 4 respectively. This analysis was repeated for thin films strained by a LAO substrate (Fig. 4(c)), producing the same results.

According to the first-principles calculations, the difference in dielectric function between a large and a small CMO supercell originates from a structural reconstruction on the surface between the supercell and the vacuum, the energy associated with the reconstruction being $\sim$3 eV for the total surface model system. This value is much larger than 0.07 eV, which is the computed energy difference between bulk G-type antiferromagnetic CMO at zero temperature and paramagnetic bulk CMO at room temperature. These are modeled by the structural parameters relaxed to the ground state and using an approximate paramagnetic spin configuration~\cite{Xiang2015}, with structural parameters taken from the experiment performed at room temperature by Bozin {\em et al.}~\cite{Bozin08}. Thus, our experimental results can be qualitatively interpreted by the first-principles calculations at 0 K assuming the G-type antiferromagnetic configuration.

The thickness dependence of $\widetilde{\epsilon}(\omega)$ correlates well with strain relaxation and surface contribution. However, other possible contributions may take place. 
For example, CMO exhibits magnetic ordering, but the N\'{e}el temperature in bulk CMO, $T_\text{N, CMO}$, is $\sim$130 K ($\ll$ 300 K)~\cite{nicastro2002exchange}, and it is highly unlikely that strain and/or finiteness of the film allows magnetic ordering at 300 K. It has been demonstrated that under high tensile strain oxygen vacancy formation in CMO is favored ~\cite{aschauer2013strain}, which could alter band energies. These changes in band energies could alter the electronic and optical properties of thin films and their contribution, although not addressed in this work, cannot be ruled out.

We have shown that $\widetilde{\epsilon}(\omega)$ decreases in magnitude for decreasing CMO film thickness, particularly as the strain transitions from partially relaxed to coherently strained. We observed that for the thinnest films this evolution of $\widetilde{\epsilon}(\omega)$ continues even under the same strain state. Taken together, our DFT results, combined with the spectroscopic ellipsometry analysis of the films, indicate that in the thinnest films the surface contribution is dominant, whereas in the thicker films progressive partial and full strain relaxation dominates the evolution of $\widetilde{\epsilon}(\omega)$. The interplay among thickness-driven large surface and substrate-induced strain contributions to $\widetilde{\epsilon}(\omega)$ in an epitaxial perovskite oxide thin film in its non-magnetic phase holds promise for a novel route to thickness-induced engineering of optical properties.  

\setlength{\tabcolsep}{12pt}
\begin{table}[H!]
\centering
\caption{Structural parameters of CMO films in this study. Thickness (\textit{t}) was determined from XRR and VASE. Thickness from XRR data was determined first and used initially during VASE parameter fitting. Film in-plane lattice parameter ($a_\parallel$) is extracted from GIXRD (for films $<$20 nm) or RSM (for films $>$20 nm) characterization and data fitting. MSE is mean squared error from VASE fitting. Surface roughness was determined from XRR and VASE. The surface roughness as determined by XRR is given by the fittings of XRR data. Fitting of the XRR data provides a roughness value being the standard deviation of the rms error function. The VASE roughness error is determined from the 90\% confidence limit of the value of the fitting parameter used to determine surface roughness. All error values are in parenthesis.}
\label{Table I.}
\begin{tabular}{@{}ccccccc@{}}
\toprule
\toprule
{\bf Substrate}      & {\bf \begin{tabular}[c]{@{}c@{}}\textit{t}, XRR\footnotemark[1]\\(nm)\end{tabular}} & {\bf \begin{tabular}[c]{@{}c@{}}\textit{t}, VASE\\(nm)\end{tabular}} & {\bf \begin{tabular}[c]{@{}c@{}}$a_\parallel$ (\AA) \end{tabular}} & {\bf \begin{tabular}[c]{@{}c@{}}In-plane\\strain\%\end{tabular}} & {\bf \begin{tabular}[c]{@{}c@{}}MSE\end{tabular}} & {\bf \begin{tabular}[c]{@{}c@{}}Film roughness,\\VASE (nm)\end{tabular}} \\ \midrule
STO(001)               & 4.1(0.18)                                                    & 4.1(0.79)                                                     & 3.905\footnotemark[2]                                                    & 5.0                & 1.52                                                          & 0.3(0.04)                                                      \\
                     & 5.8(0.14)                                                   & 6.1(0.01)                                                     & 3.905\footnotemark[2]                                                    & 5.0               & 2.08                                                          & 0(0.02)                                                           \\
                     & 7.1(0.19)                                                   & 7.0(0.19)                                                     & 3.90(0.002)                                                                              & 4.8               & 1.93                                                          & 0(0.01)                                                             \\
                     & 10.4(0.19)                                                  & 10.3(0.97)                                                    & 3.76(0.002)                                                                              & 1.1               & 3.61                                                          & 0(0.05)                                                              \\
                     & 62.9(0.17)                                                  & 60.8(0.18)                                                   & 3.75(0.001)                                                                              & 0.8               & 2.71                                                         & 0(0.12)                                                              \\
LAO(001)               & 9.6(0.85)                                                  & 9.6(0.21)                                                      & 3.790\footnotemark[2]                                                    & 1.9               & 4.07                                                           & 0.28(0.05)                                                    \\
                     & 4.9(0.24)                                                 & 4.3(0.07)                                                      & 3.790\footnotemark[2]                                                     & 1.9              & 2.99                                                           & 0(0.02)                                                     \\
SLAO(001)                 & 9.0(0.09)                                               & 10.0(0.94)                                                   & 3.754\footnotemark[2]                                                     & 0.91               & 5.16                                                           & 0.1(0.05)                                                     \\
\multicolumn{1}{l}{} & 4.3(0.37)                                        & 4.2(0.04)                                                    & 3.754\footnotemark[2]                                                    & 0.91               & 4.73                                                           & 0(0.04)                                             \\ \bottomrule \bottomrule
\vspace{-25pt}
\footnotetext[1] {XRR thickness error is the surface roughness.}
\footnotetext[2]{GIXRD data indicates that the film is fully strained to the substrate. The lattice parameter given is that for the substrate.}
\end{tabular}
\end{table}

\begin{figure}[H!]
\includegraphics[scale=0.6]{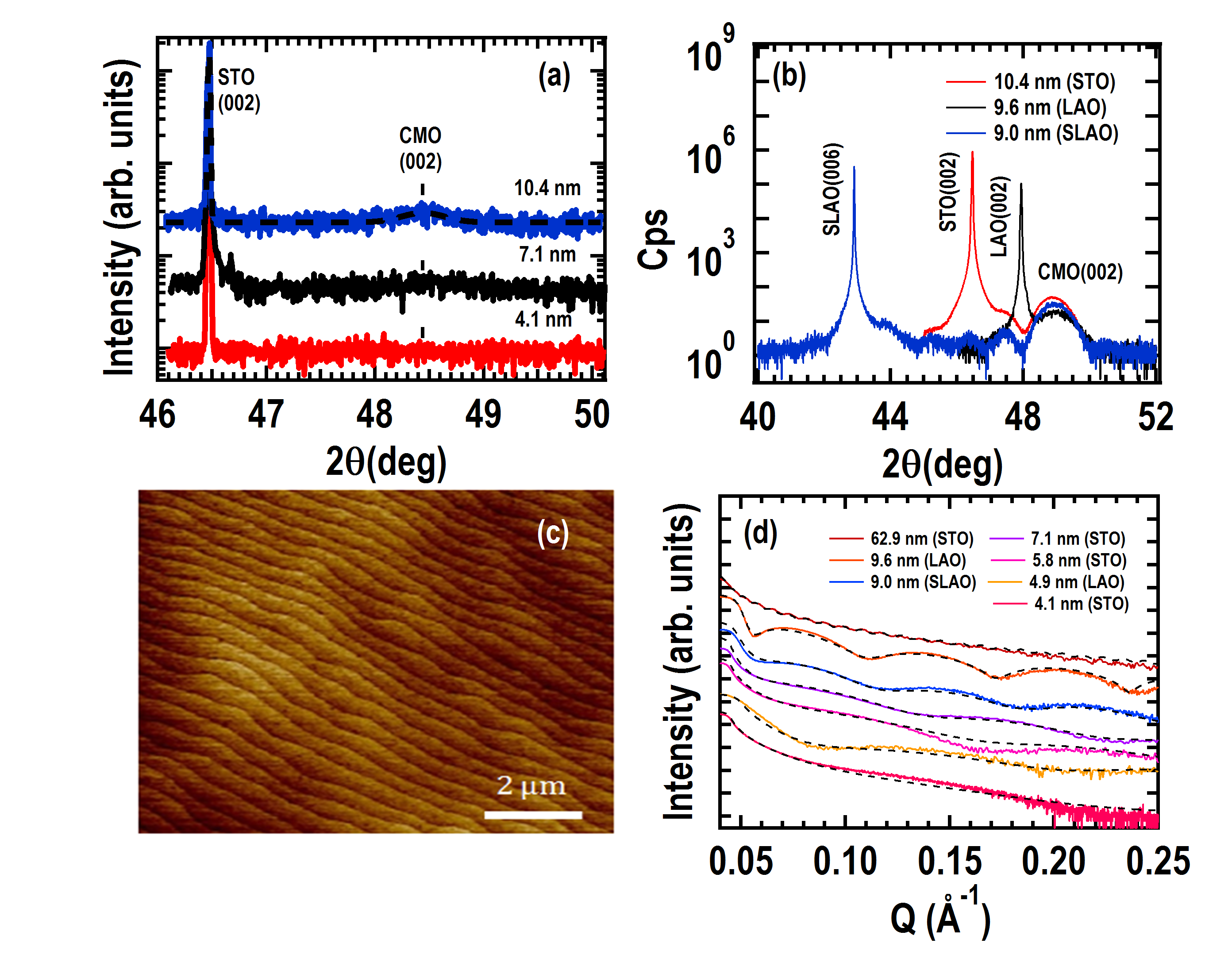}
\caption{(\textit{a}) GIXRD, (\textit{b}) XRD,(\textit{c}) AFM and (\textit{d}) XRR data of CMO films with varying thickness grown on different substrates. (\textit{a}) GIXRD spectra of a 10.4, 7.1, and 4.1 nm thick CMO film grown on STO substrates. The film peak is at 48.44$^\circ$. Fitting is shown (dashed line) for the 10.4 nm film. (\textit{b}) The 2$\theta/\omega$ XRD scan of CMO thin films ($\sim$10 nm in thickness) on each substrate. Substrate peaks and film peaks (48.93$^\circ$) are identified. (\textit{c}) AFM of the 10.4 nm thick CMO film on STO showing the underlying terraces from the annealed STO substrate, indicating an atomically flat film. (\textit{d}) The XRR spectra for films with different thickness on STO, LAO and SLAO substrates. Fitting is given for each trace (dashed line).}
\label{fig:FIG_1}
\end{figure}

\begin{figure}[H!]
\includegraphics[scale=0.6]{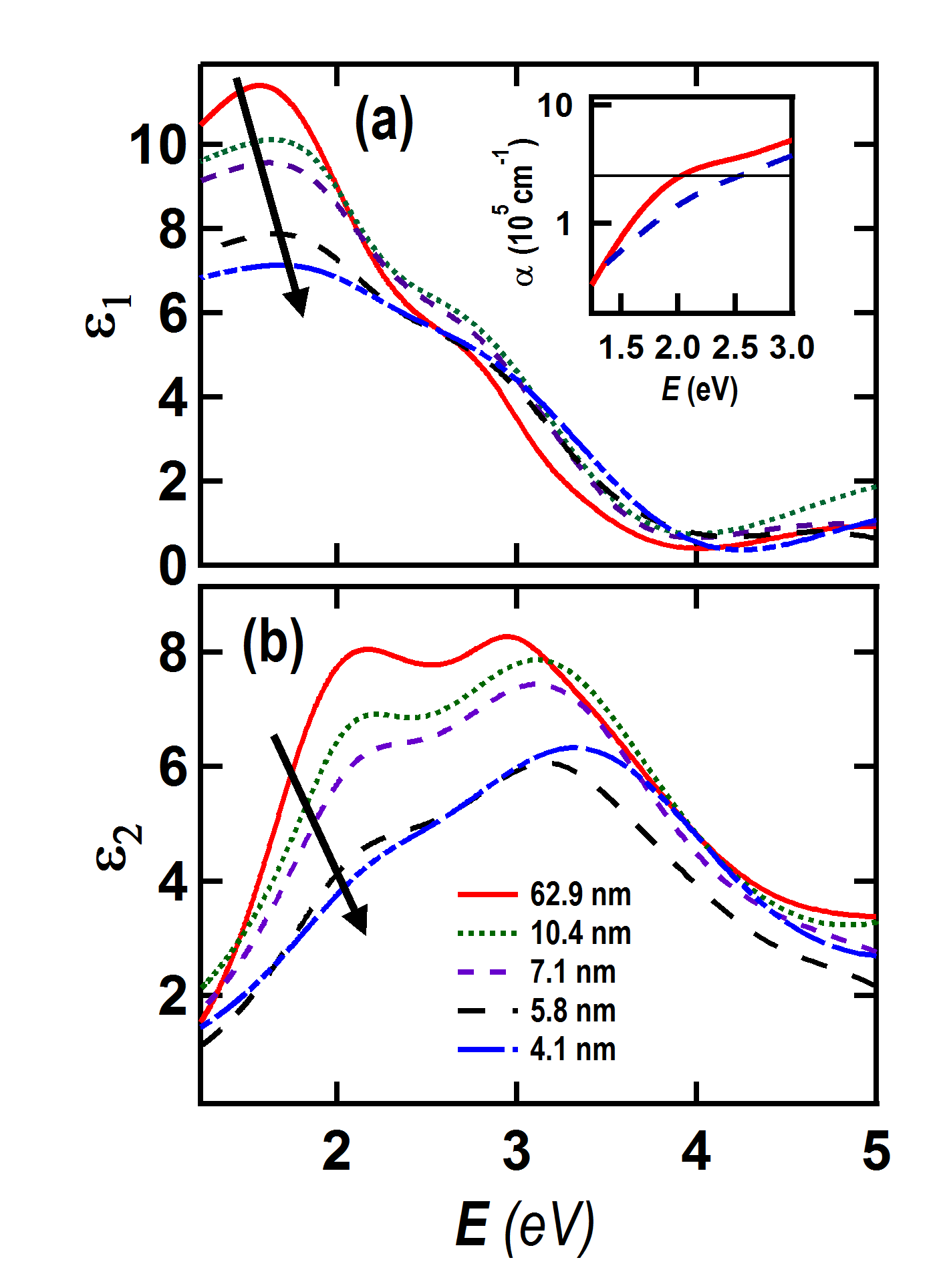}
\caption{(\textit{a}) real, $\epsilon_{1}$, and (\textit{b}) imaginary part, $\epsilon_{2}$, of the experimentally determined complex frequency-dependent dielectric function as a function of photon energy for different CMO film thicknesses on STO(001). The arrow denotes films of decreasing thickness, as specified in Table I. The thinnest two films are strain coherent, as the films increase in thickness they begin to strain relax. The inset shows $\alpha$ for the 62.9 nm (solid) and 4.1 nm (dotted) films plotted on a semi-log scale over the range 1.2 - 3 eV. At a fixed value of absorption (solid line shows $\alpha$ = 2.5$\times$10$^5$ cm$^{-1}$) there is as much as a $\sim$0.5 eV spectral shift in the photon energy for decreasing film thickness.}
\label{fig:FIG_2}
\end{figure}

\begin{figure}[H!]
\includegraphics[scale=0.6]{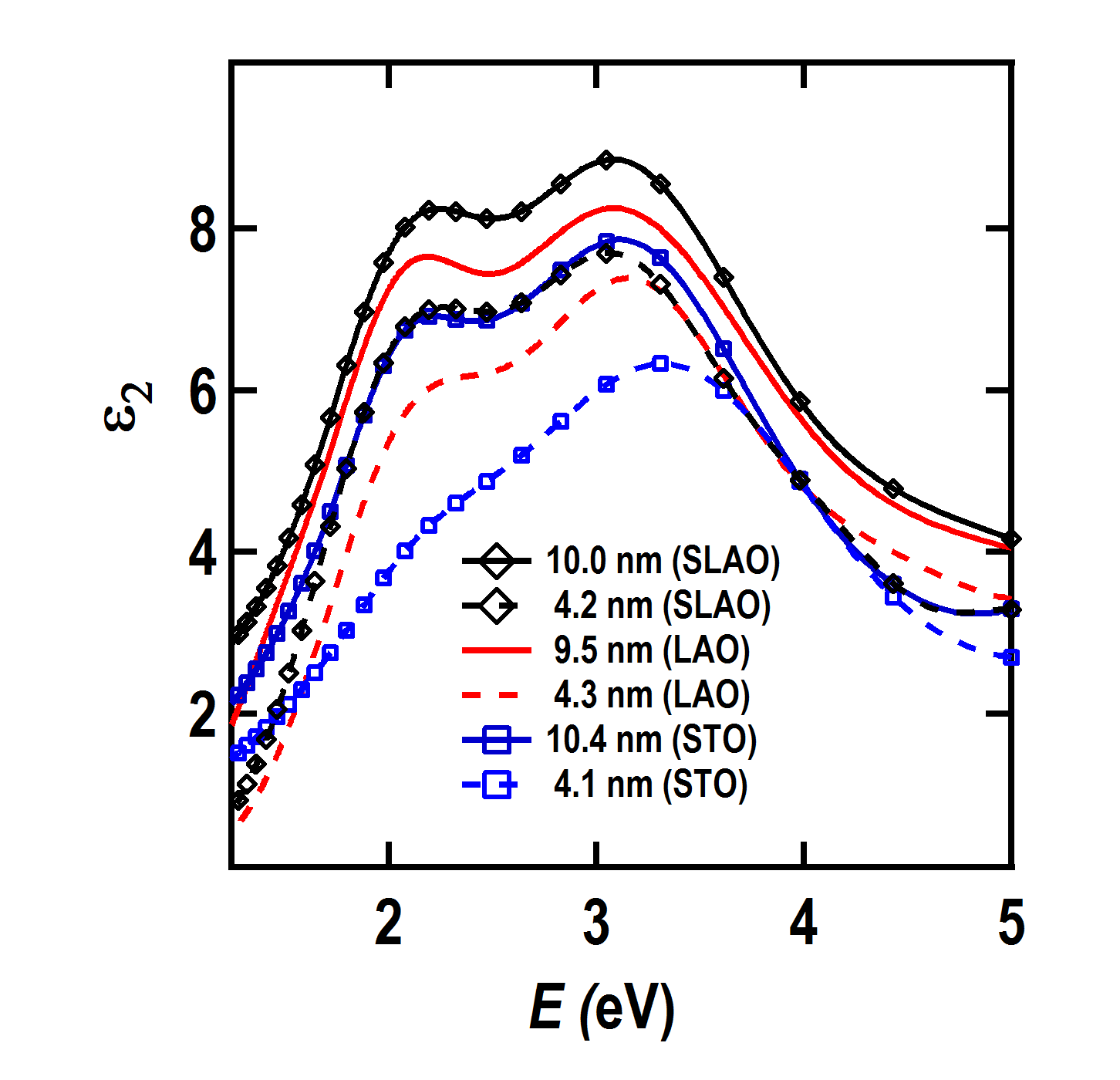}
\caption{Imaginary part, $\epsilon_{2}$, of the experimentally determined complex frequency-dependent dielectric function as a function of photon energy for different CMO film thicknesses on three different substrates. The legend in the figure is the CMO film thickness on the indicated substrate. The dielectric function of the CMO/SLAO film (diamond marker) is closest of that of bulk CMO as compared to CMO grown on LAO (no marker) or STO (square marker). Thicker films are shown with a solid line, the thinner films are shown with a dashed line.}
\label{fig:FIG_3}
\end{figure}

\begin{figure}[H!]
  \includegraphics[scale=0.6]{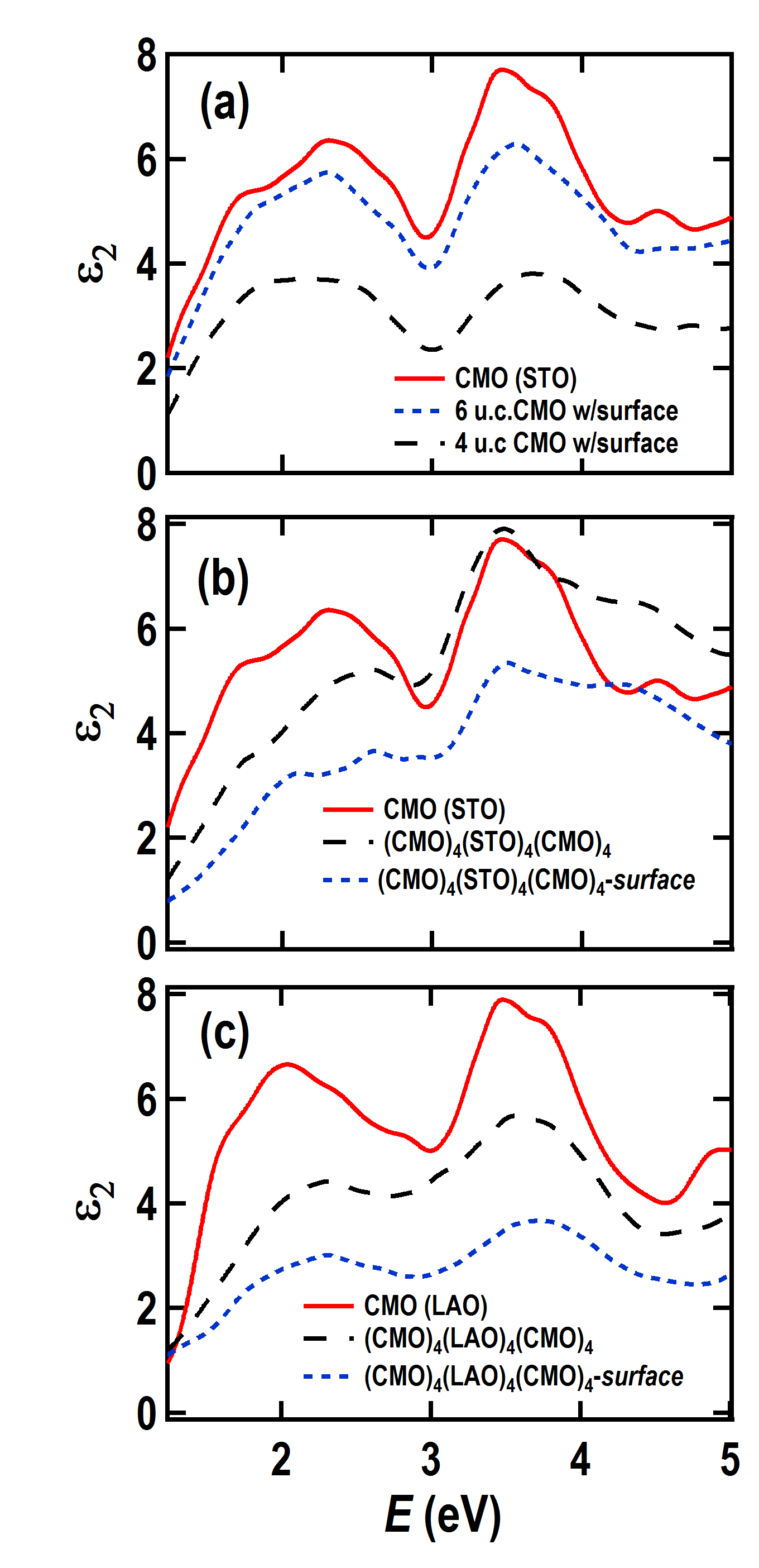}
  \caption{(\textit{a}) The effect of free energy-minimized surface termination on the imaginary part of the complex dielectric function ($\epsilon_{2}$) in the in-plane lattice direction for four- and six-unit cell structures with a free surface as compared to that for bulk CMO strain coherent with STO (CMO(STO)) or LAO (CMO (LAO)). (\textit{b}) $\epsilon_{2}$ in the in-plane lattice direction incorporating epitaxial strain by introducing STO into the supercell both with no free surface ((CMO)$_4$(STO)$_4$(CMO)$_4$) and one free surface ((CMO)$_4$(STO)$_4$(CMO)$_4$-${surface}$). (\textit{c}) $\epsilon_{2}$ in the in-plane lattice direction incorporating epitaxial strain by introducing LAO into the supercell both with no free surface ((CMO)$_4$(LAO)$_4$(CMO)$_4$) and one free surface ((CMO)$_4$(LAO)$_4$(CMO)$_4$-${surface}$).}
  \label{fig:FIG_4}
\end{figure}

\subsection{Acknowledgments}
The authors acknowledge the Air Force Office of Scientific Research under FA9550-13-1-0124. J.E.S. also acknowledges the U.S. Army Research Office for support of A.R.A. under W911NF-14-1-0500. J.E.S. thanks C. L. Schauer for access to the spectroscopic ellipsometer. 

\pagebreak
\bibliography{bib_file}

\end{document}